\documentclass[fleqn,usenatbib]{mnras}

\usepackage{newtxtext,newtxmath}

\usepackage[T1]{fontenc}

\DeclareRobustCommand{\VAN}[3]{#2}
\let\VANthebibliography\thebibliography
\def\thebibliography{\DeclareRobustCommand{\VAN}[3]{##3}\VANthebibliography}

\usepackage{graphicx}	
\usepackage{amsmath}	
\usepackage{xcolor}

\defcitealias{cameron2019}{C+19}

\title[A novel analysis of LBGs contamination]{A novel analysis of contamination in Lyman-break galaxy samples at $\boldsymbol{z\sim6-8}$: spatial correlation with intermediate-redshift galaxies at $\boldsymbol{z\sim1.3-2}$}

\author[M. Hilmi et al.]{
Miftahul Hilmi,$^{1,2}$\thanks{E-mail: mhilmi@student.unimelb.edu.au}
Nicha Leethochawalit,$^{3,1,2}$
Michele Trenti$^{1,2}$
and Benjamin Metha$^{1,2}$
\\
$^{1}$School of Physics, the University of Melbourne, VIC 3010, Australia\\
$^{2}$ARC Centre of Excellence for All Sky Astrophysics in 3 Dimensions (ASTRO 3D), Australia\\
$^{3}$National Astronomical Research Institute of Thailand (NARIT), Mae Rim, Chiang Mai, 50180, Thailand
}

\date{Accepted XXX. Received YYY; in original form ZZZ}

\pubyear{2023}

\begin{document}
\label{firstpage}
\pagerange{\pageref{firstpage}--\pageref{lastpage}}
\maketitle

\begin{abstract}
Potential contamination from low/intermediate-redshift galaxies, such as objects with a prominent Balmer break, affects the photometric selection of high-redshift galaxies through identification of a Lyman break. Traditionally, contamination is estimated from spectroscopic follow-up and/or simulations. Here, we introduce a novel approach to estimating contamination for Lyman-break galaxy (LBG) samples based on measuring spatial correlation with the parent population of lower redshift interlopers. We propose two conceptual approaches applicable to different survey strategies: a single large contiguous field and a survey consisting of multiple independent lines of sight. 
For a large single field, we compute the cross-correlation function between galaxies at redshift $z \sim 6$ and intermediate-redshift galaxies at $z \sim 1.3$. We apply the method to the CANDELS GOODS-S and XDF surveys and compare the measurement with simulated mock observations, finding that the contamination level in both cases is not measurable and lies below $5.5\%$ (at 90\% confidence). 
For random-pointing multiple field surveys, we measure instead the number count correlation between high-redshift galaxies and interlopers, as a two-point correlation analysis is not generally feasible. We show an application to the LBG samples at redshift $z \sim 8$ and the possible interloper population at $z \sim 2$ in the Brightest of Reionizing Galaxies (BoRG) survey. By comparing the Pearson correlation coefficient with the result from Monte Carlo simulations, we estimate a contamination fraction of $62^{+13}_{-39}\%$, consistent with previous estimates in the literature.
These results validate the proposed approach and demonstrate its utility as an independent check of contamination in photometrically selected samples of high-redshift galaxies. 


\end{abstract}




\section{Introduction}




Thanks to space-based observatories, thousands of galaxy candidates at redshift $z\gtrsim6$ have been recently discovered, primarily from several large survey programs conducted with the Hubble Space Telescope \citep{schmidt2014, bouwens2015, ishigaki2015, morishita2018, bowler2020, salmon2020, roberts-borsani2022}.
The identification of high-redshift galaxy candidates is conducted photometrically through the Lyman-break technique \citep{steidel1996}, which relies on the identification of a strong spectral break at a wavelength shorter than 1216 \AA.
This method heavily depends on the color information of the sources and is therefore subject to contamination from objects with similar photometry, such as cool stars or intermediate-redshift red galaxies. In particular, one of the main sources of contamination for Lyman-break galaxy (LBG) samples are low/intermediate redshift Balmer break galaxies with a prominent break at $3646$ \AA{} rest frame \citep{atek2011, vanderWel2011}.  

To minimize contamination in photometric catalogs, deep observations at wavelengths shorter than the spectral break are generally required to distinguish between a faint continuum of an interloper and a true non-detection for a high-redshift galaxy. For example, \cite{stanway2008} suggest using a set of non-overlapping but adjacent filters to be able to impose a clear color cut on the selection and thus reduce contamination. 
Deeper imaging follow-up observations on previously identified candidates also shows that additional photometry blueward of the Lyman break can help discriminating between low and high-redshift galaxies (e.g., \citealt{livermore2018}). Yet, contamination is unavoidable in photometrically selected samples, and thus needs to be understood. 

Contamination from intermediate-redshift galaxies can contribute to bias in estimating the high-redshift UV Luminosity Functions \citep{morishita2018}, in addition to other sources of bias, such as magnification bias \citep{wyithe2011, mason2015}, bias due to the cosmic variance \citep{trenti2008, moutard2016, bowler2020}, and bias due to photometric scatter \citep{Leethochawalit2022a}.
Previous studies also show that contamination levels becomes higher with increasing redshift. \cite{vulcani2017} found that the ratio of interlopers to dropouts grows significantly as a function of redshift. 
Using a simple model that relies on the dark matter halo mass function, \cite{furlanetto2023} also found that the expected contamination level increases drastically at $z\gtrsim10$, requiring stricter selection criteria to identify high-redshift sources robustly.

While spectroscopic observations are the most robust approach to verify high-redshift candidates, they require large investment of telescope time and/or are unfeasible for objects near the detection limit of imaging surveys.  Thus, several works have proposed methods to estimate the contamination level.
\cite{finkelstein2015} artificially dim real lower redshift sources to see if this allows them to be selected as high-redshift candidates. This method implicitly assumes that contaminants have similar spectral energy distributions (SEDs) to the known lower redshift sources. They estimate a relatively small contamination fraction of $\sim5\%-15\%$ in the CANDELS GOODS fields, which is in agreement with the estimation from the candidates' redshift probability distributions produced by photometric redshift fitting code ($P(z)$ curves).
In another work, \citet{rojas-ruiz2020} estimate the contamination by downgrading deep images from the Hubble Frontier Fields program \citep{lotz2017} to the depth of the shallower Hubble images used in their work. 
By comparing the redshifts determined across the six HFF fields and the redshifts determined in the counterpart downgraded images, they found one contaminant and concluded that contaminants do not contribute significantly to their sample. Alternatively, \citet{trenti2011} apply the color selection used to select their high redshift sample to a library of SED models of lower redshift galaxies taking into account the depth of the observations for the survey modelled.

This paper proposes a novel conceptual framework to assess contamination, and presents two implementations of the idea. The first approach is based on the spatial correlation between the high-redshift galaxy candidates and known galaxies at the redshift of potential interlopers. It is appropriate for a large contiguous survey. The basic principle is that the angular cross-correlation function of high-redshift and intermediate-redshift galaxies should not indicate any clustering, unless some level of contamination exist. 
This approach is inspired by \citet{menard2013,schmidt2014,rahman2016a,rahman2016b}, where clustering analyses were proposed to refine photometric redshift estimates. These works typically consider two populations: a reference population with known redshift and angular positions, and the other population with only angular positions known. The redshifts of the second population can be determined when there is a cross-correlation signal with the reference population.
The concept of spatial correlation has also been applied in other studies to measure contamination in various samples. \cite{gebhardt2019} and \cite{farrow2021} use the cross-correlation function to estimate the contamination fraction of low redshift [OII] ($z<0.5$) emitters in the intermediate redshift Ly$\alpha$ emitters sample ($1.9<z<3.5$) to \textcolor{red}{estimate} the unbiased cosmological parameters. 
\cite{addison2019} also suggests the use of cross-correlation function to constrain the contamination fraction in [OIII] sources sample due to the misidentification of H$\alpha$ spectral line. 
\citet{awan2020} presents a correlation function estimator that can correct for sample contamination, by taking into account the auto and cross-correlation function of the sources and contaminants.
In our work, we take this concept to study contamination of LBG samples at high redshift.

The second approach is to quantify the number count correlation between high-redshift galaxies and the possible contaminants at intermediate redshift. It is appropriate for analysing the contamination level of a random-pointing survey with multiple fields. The approach is adapted from the counts-in-cells method proposed by \cite{robertson2010} to quantify the clustering properties of galaxies for observations that consist of a large number of uncorrelated fields, which has been implemented by \citet{cameron2019} on BoRG observations. These methods are based on the sources’ angular positions and number counts. They therefore minimize the reliance on manipulating/analyzing the SEDs of candidates and on simulated high-redshift galaxies, and provide an independent way to cross-check estimates obtained through traditional methods.

This paper is organized as follows. 
In Section 2, we describe the angular cross-correlation technique to estimate contamination fractions and apply it to CANDELS data. 
In Section 3, we model the cross-correlation function using mock catalogs generated from IllustrisTNG simulation, to determine what level of contamination this technique is sensitive to. 
Section 4 discusses the number count analysis based on BoRG samples. 
We summarize our results and conclusion in Section 5.
Throughout the paper, we adopt a cosmological parameter set of $\Omega_\mathrm{M} = 0.3$, $\Omega_\Lambda = 0.7$, and $H_0 = 70\ \mathrm{km\ s^{-1} Mpc^{-1}}$. 
All magnitudes are represented in the AB system \citep{oke1983}.

\section{Cross-Correlation Analysis}\label{sec:cross-correlation}


This Section explores the spatial correlation between high-redshift galaxies and lower-redshift galaxies at the \textit{interloper redshift}, which we define to be the redshift range in which intermediate-redshift galaxies resemble high-redshift galaxies photometrically, and may contaminate the high-redshift galaxies samples. More specifically, this is the redshift where the observed Balmer break of intermediate-redshift galaxies is at the same wavelength as the observed Lyman break of high-redshift galaxies:
\begin{equation}
    1216\ \text{\r{A}} (1 + z_{\text{high}}) = 3646\ \text{\r{A}} (1 + z_{\text{interloper}}).
\end{equation}
Here, the Ly$\alpha$ wavelength $1216\ \text{\r{A}}$ is used instead of the Lyman limit $912\ \text{\r{A}}$, because for high-redshift galaxies at $z\gtrsim6$, the continuum between $912\ \text{\r{A}}$ and $1216\ \text{\r{A}}$ is absorbed by intervening Ly$\alpha$ forest \citep{madau1995,giavalisco2002}.

The method rests on the lack of physical correlation between galaxies at high redshift and galaxies at interloper redshift since the typical correlation length of dark-matter halos is orders of magnitude smaller than the comoving line-of-sight distance between the two populations. Therefore, the two samples should be uncorrelated unless some galaxies at the lower redshift are misidentified as high-redshift galaxies and contaminate the high-$z$ sample. Based on this, we hypothesize that we should be able to constrain the contamination rate based on the spatial correlation between high-redshift candidates and known galaxies at the interloper redshift. The so-called Schrodinger's galaxy presented in \citet{naidu2022} is a good illustration of this idea. The SED fitting of the galaxy suggests that the galaxy is at $z\sim17$ with a small probability to be at $z\sim5$. However, the galaxy is in the vicinity of three neighbouring galaxies that are at $z\sim5$. Hence, the authors suggest that the source could also likely be part of the protocluster. 

\subsection{Data Set}
To obtain statistically robust spatial correlations, we need large samples of galaxies at both high redshift and interloper redshift observed in the same survey with a large contiguous area. With this requirement, we use the data set of the GOODS-South and the XDF fields from the Hubble Legacy Fields Data Release V2.5 \citep{illingworth2016,whitaker2019}\footnote{\href{https://archive.stsci.edu/prepds/hlf/}{https://archive.stsci.edu/prepds/hlf/}}. The area of the GOODS-S survey is 64.5 arcmin\textsuperscript{2}, while for XDF it is 4.7 arcmin\textsuperscript{2}. We decided to use $z\sim 6$ (specifically $z=5.5 - 6.5$) galaxies as our main high-$z$ sample to ensure the sample is sufficiently large to enable two-point correlation function measurements. The corresponding interloper redshift is $z=1.2 - 1.5$ with the average redshift equal to $z\sim1.3$. We use the catalog from \citet{merlin2021} to obtain the sample of intermediate-redshift galaxies, and the catalog from \citet{bouwens2021} as the high-redshift galaxies sample. \citet{merlin2021} selected their samples in $H_{160}$ band and used SED fitting to determine the redshifts for objects with no spectroscopic redshifts available. On the other hand, the $z\sim6$ samples in \citet{bouwens2021} are detected in $Y_{105}J_{125}JH_{140}H_{160}$ stacked images and are selected based on Lyman-break color criteria.

\begin{figure}
    \centering
    \includegraphics[width=0.46\textwidth]{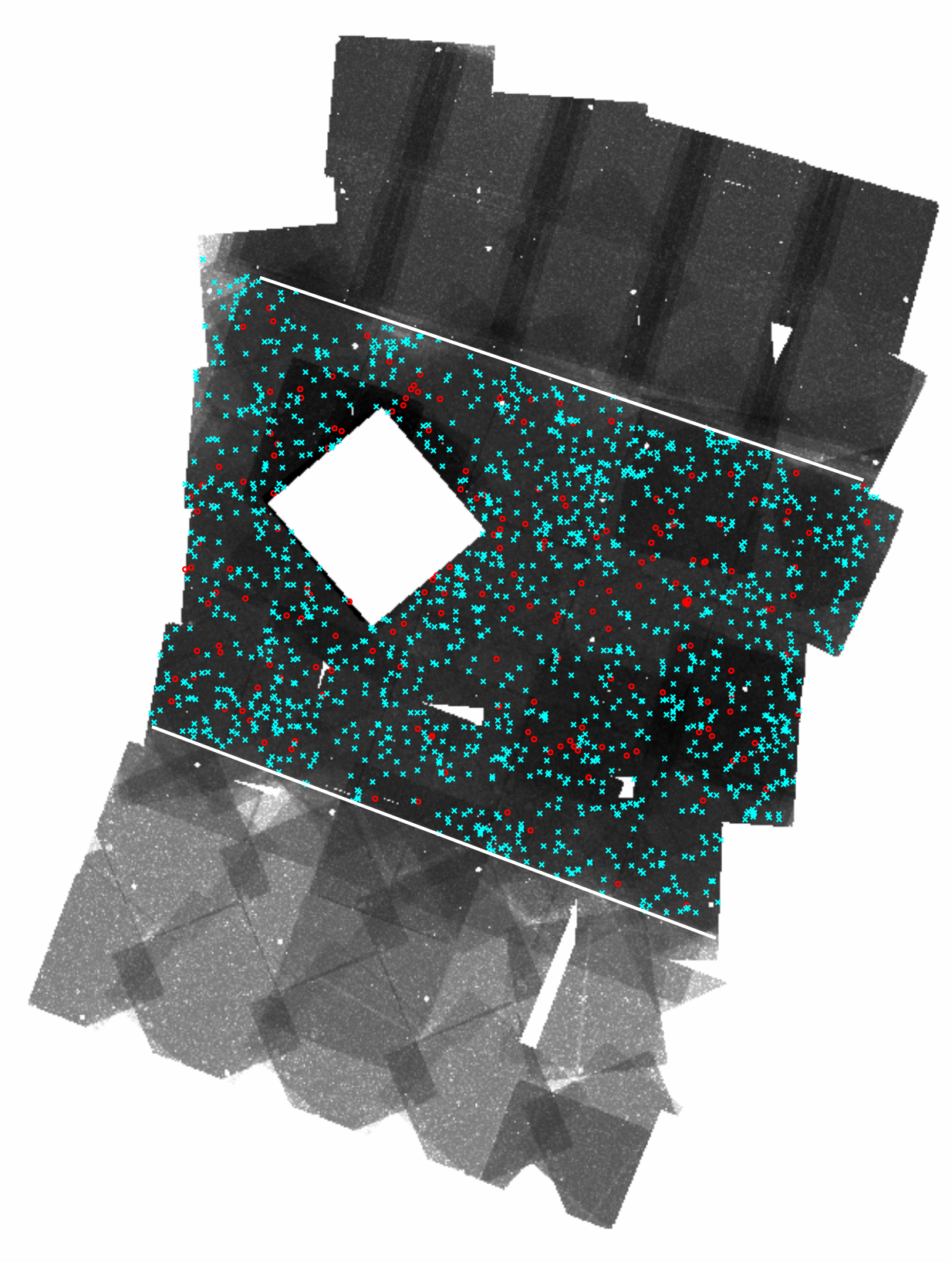}
    \caption{Root mean square (rms) image of GOODS-S field taken in F125W band and shown in logarithmic scale. As we can see from the image, the central region has a different depth compared to the edge regions. Therefore, we only consider galaxies inside the white lines, with sources marked as cyan crosses for $z\sim1.3$ galaxies and red circles for $z\sim6$ galaxies.}
    \label{fig:goods_image}
\end{figure}

Based on the catalog from \citet{merlin2021}, we select galaxies at redshift $z=1.2$ to $z=1.5$, yielding 3379 and 292 galaxies located on GOODS-S and XDF fields, respectively. For high-$z$ galaxies, we select the galaxies at redshift $5.6<z<6.5$ from \citet{bouwens2021}. There are 323 and 129 such galaxies on GOODS-S and XDF fields, respectively.
Due to the different depths between the edge part and central part of the GOODS-S survey, the completeness of the survey is non-homogeneous and this may introduce systematic errors in our cross-correlation analysis.
Therefore, for the GOODS-S area, we restrict the samples to those in the central region with a uniform depth (see Figure \ref{fig:goods_image}). 

As the two catalogs are provided by different studies, we investigated if there is any common candidate in the $z\sim1.3$ and $z\sim6$ catalogs. 
We find that there are 8 sources in the GOODS-S field and 3 sources in the XDF field that are reported in both catalogs. 
We removed those sources from $z\sim1.3$ catalog and assigned them only to the high-redshift catalog, since our study aims to check the quality of the $z\sim6$ catalog.
Our final samples consist of 1387 $z\sim1.3$ galaxies and 191 $z\sim6$ galaxies in the GOODS-S field. The locations of the samples are also shown in Figure \ref{fig:goods_image}. 
For XDF, our final samples contain 289 and 129 sources in the $z\sim1.3$ and $z\sim6$ catalog, respectively.
As an additional note, one of the $z\sim6$ galaxies in GOODS-S sample is also spectroscopically confirmed to be at $z\sim1.3$ \citep{vanzella2008}, but the source is retained in the photometric sample. 
While it is possible that high-redshift galaxies are misidentified as intermediate-redshift galaxies, the fraction will be very small. The number of galaxies at intermediate-redshift is much higher than those at high-redshift. Therefore, the contamination in intermediate-redshift galaxies sample by high-redshift galaxies is assumed to be zero.

\subsection{Analysis}
\label{subsec:cross_corr_obs}

The angular correlation function, $\omega_\textrm{cor}(\theta)$, measures the clustering of galaxies by comparing the observed number of galaxy pairs relative to the expected number of galaxy pairs from a random distribution. The angular correlation function of galaxies at any redshift can generally be described by a power law function: $\omega_\textrm{cor} = A_\mathrm{\omega}\theta_i^{-\beta}$ \citep{lee2006,overzier2006,barone-nugent2014}.

\begin{figure*}
    \centering
	\includegraphics[width=0.9\textwidth]{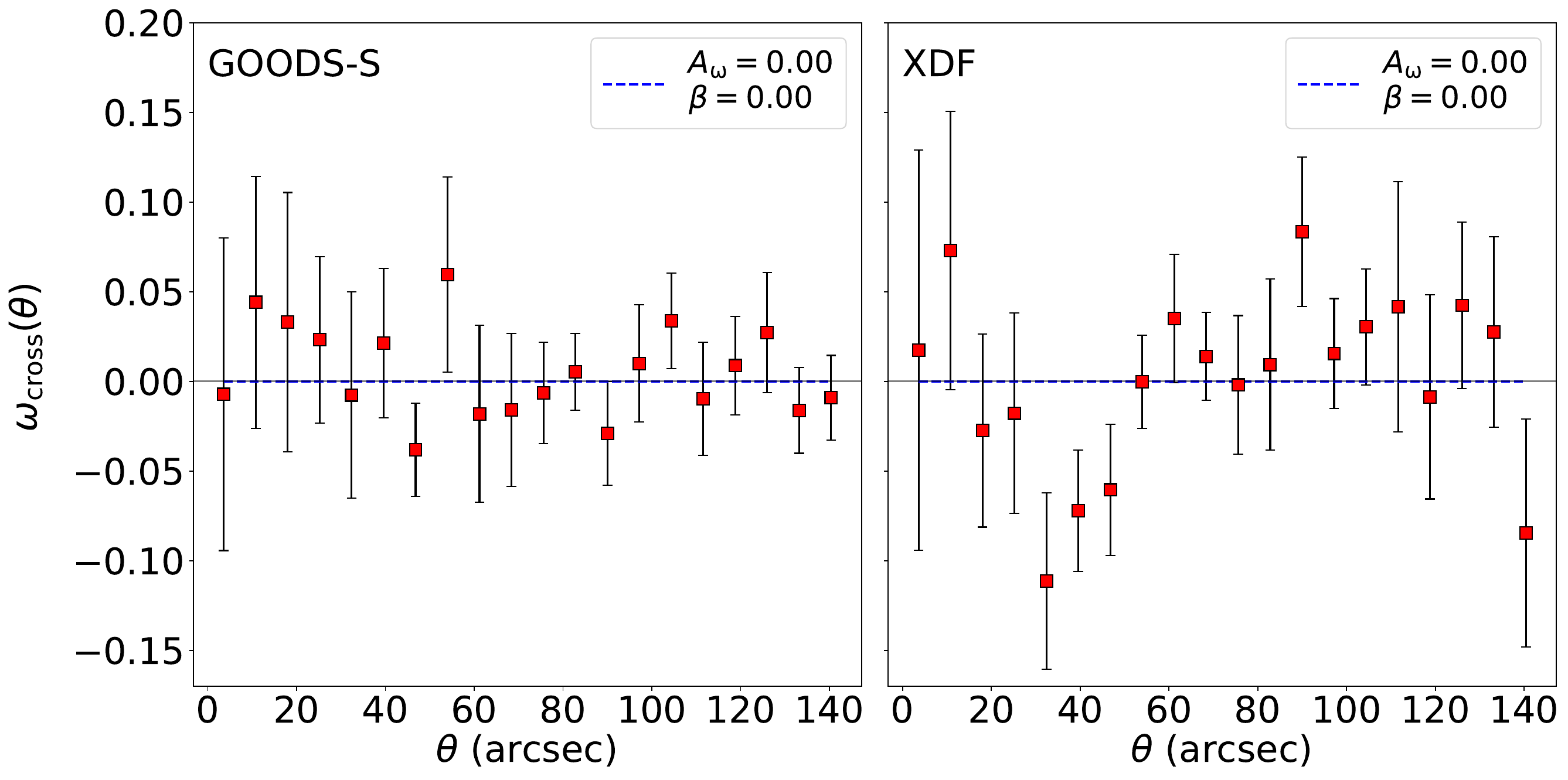}
    \caption{Cross-correlation function of $z\sim1.3$ and $z\sim6$ galaxies in the GOODS-S field (left) and XDF (right). Dashed line is the power law function of $\omega_\mathrm{cross}(\theta)=A_\mathrm{\omega}\theta^{-\beta}$, where $A_\mathrm{\omega}$ and $\beta$ are the best-fit parameters constrained by the $\chi^2$ fitting with Equation \ref{eq:model1}.}
    \label{fig:cross-corr}
\end{figure*}

To analyze whether there is significant contamination within $z\sim6$ galaxies sample, we calculate the cross-correlation function between $z\sim1.3$ and $z\sim6$ galaxies.
Similar to the angular correlation function, the cross-correlation function measures the excess probability of finding a pair of galaxies from two different populations within an angular separation $\theta$. 
We use the modified Landy-Szalay estimator \citep{landy1993,blake2006}:
\begin{equation}
    \omega_\mathrm{cross}(\theta)=\frac{D_1D_2(\theta)-D_1R_2(\theta)-D_2R_1(\theta)+R_1R_2(\theta)}{R_1R_2(\theta)},
	\label{eq:cross-corr}
\end{equation}
where $D_1D_2(\theta)$, $D_1R_2(\theta)$, $D_2R_1(\theta)$, and $R_1R_2(\theta)$ are the number of $z\sim1.3$ galaxy and $z\sim6$ galaxy pairs, $z\sim1.3$ galaxy and random point pairs, $z\sim6$ galaxy and random point pairs, and random-random point pairs, all measured within an angular separation of $\theta\pm\delta_\theta$, respectively. 
For our study, the random point catalog is generated by taking the depths of fields in each filter into account in the same manner as described in details in \citet{Dalmasso2023}. This process is undertaken to prevent artificial clustering signals induced by non-uniform depth variations. In summary, we randomly inject galaxies with S\'ersic light profile in the images of all detection bands. The final random catalog consists of the injected galaxies that are recovered with the same procedures used for galaxy detection in the GOODS-S \citep{merlin2021} and XDF \citep{bouwens2021} catalogs. For simplicity, we use the same random catalog for both galaxy samples ($R_1=R_2=R$).
We estimate the cross-correlation function in bins of $\theta$, using linear binning with a bin width of $\delta_\theta=7\farcs2$.
We use Bootstrap resampling \citep{ling1986} to estimate errors in the cross-correlation function by resampling the dataset ten times. We show the resulting cross correlation functions with red squares and black error bars in Figure \ref{fig:cross-corr}.
Visually, there is no correlation signal in both GOODS-S (left panel) and XDF (right panel) fields.

To investigate the significance of the cross-correlation signal, we conduct a statistical test between the two models of the cross-correlation function. 
If there is significant contamination, we expect a clustering signal between galaxies at $z\sim1.3$ and $z\sim6$. The intrinsic correlation functions of galaxies at both redshifts are power-law functions. Since both populations 
have similar power-law slopes \footnote{Studies of galaxy correlation function often assume a fixed power-law slope. For example, $\beta=0.8$ is assumed for all galaxies across $z=0-6$ in \citet{arnouts1999}. More recent works \citep[e.g.,][]{barone-nugent2014,Dalmasso2023} use a fixed $\beta=0.6$ for $z\gtrsim4$ galaxies. We measured the correlation function for our $z\sim1.3$ galaxies (see Section \ref{sec:f_cont_awan}). The measured $\beta$ values are $0.45\pm0.19$ and $0.81\pm0.79$ for the GOODS-S and the XDF fields, respectively. They are consistent with the power-law slopes assumed for $z=6$ galaxies in the literature.}, we therefore expect the cross-correlation function of these two populations also to take the functional form of a power-law function:
\begin{equation}
    \omega_\mathrm{cross}(\theta)=A_\mathrm{\omega}\theta^{-\beta}
    \label{eq:model0}
\end{equation}

However, due to the finite survey area, observed cross-correlation function $\omega_\mathrm{cross, obs}(\theta)$ are generally underestimated by a constant factor known as \textit{integral constraint} (IC):
\begin{equation}
    \omega_\mathrm{cross, obs}(\theta)=A_\mathrm{\omega}\theta^{-\beta}-\mathrm{IC}(A_\mathrm{\omega},\beta).
    \label{eq:model1}
\end{equation}
IC can be calculated by doubly integrating the cross-correlation function $\omega_\mathrm{cross}(\theta)$ over the survey area $\Omega$ \citep{roche1999}:
\begin{equation}
\begin{aligned}
    \mathrm{IC} & = \frac{1}{\Omega^2}\int_1\int_2\omega_\mathrm{cross}(\theta)d\Omega_1d\Omega_2\\
    & = \frac{\Sigma_iRR(\theta_i)\omega_\mathrm{cross}(\theta_i)}{\Sigma_iRR(\theta_i)}=\frac{\Sigma_iRR(\theta_i)A_\mathrm{\omega}\theta_i^{-\beta}}{\Sigma_iRR(\theta_i)}.
	\label{eq:IC}
\end{aligned}
\end{equation}
On the other hand, if there is no significant contamination, we expect that the cross-correlation function between galaxies at $z\sim1.3$ and $z\sim2$ will follow a random distribution and be given by:
\begin{equation}
    \omega_\mathrm{cross}(\theta)=\omega_\mathrm{cross, obs}(\theta)=0.
    \label{eq:model2}
\end{equation}

To take into account the correlation between measurement in different angular bins, we construct the normalized covariance matrix using the standard estimator:
\begin{equation}
    C_{ij} = \frac{1}{N-1}\sum^{N}_{l=1}\left[\omega^l(\theta_i)-\overline{\omega}(\theta_i)\right] \left[\omega^l(\theta_j)-\overline{\omega}(\theta_j)\right].
	\label{eq:cov_matrix}
\end{equation}
In this equation, the summation is over $N$ independent realizations. 
However, our Bootstrap samples are not from independent realizations. When the covariance matrix is estimated from the data itself, such as Bootstrap resampling, a correction factor of $(N-1)^2/N$ has to be added, and the covariance matrix becomes:
\begin{equation}
    C_{ij} = \frac{N_\mathrm{boot}-1}{N_\mathrm{boot}}\sum^{N_\mathrm{boot}}_{l=1}\left[\omega^l(\theta_i)-\overline{\omega}(\theta_i)\right] \left[\omega^l(\theta_j)-\overline{\omega}(\theta_j)\right],
	\label{eq:cov_matrix_boot}
\end{equation}
where $N_\mathrm{boot}$ is the total number of Bootstrap samples, $\omega^l(\theta)$ is the measured cross-correlation function from each Bootstrap realization, and $\overline{\omega}(\theta)$ is the mean of cross-correlation function. Due to the relatively small sample size, our resulting covariance matrix is noisy and the inverse of the covariance matrix is ill-conditioned and numerically unstable. Therefore, we apply a ridge regression technique \citep{hoerl1970,matthews2012} by adding a small value $c$ to the diagonal elements of the covariance matrix to reduce the impact of noise in the off-diagonal elements. We use $c=0.0001$ as our parameter value (approximately 1\% of the median value of the diagonal elements).

Using Equation \ref{eq:model1} and Equation \ref{eq:IC} together, we can estimate the best-fit parameters $A_\mathrm{\omega}$ and $\beta$ using  the $\chi^2$ minimization method under the conditions that $A_\mathrm{\omega}\ge 0$ and $\beta\ge 0$:
\begin{equation}
    \chi^2 = \sum_{i,j}\left[\omega(\theta_i)-\omega_\mathrm{model}(\theta_i)\right]^\mathrm{T} C_{ij}^{-1} \left[\omega(\theta_j)-\omega_\mathrm{model}(\theta_j)\right],
	\label{eq:chi-square}
\end{equation}
where $\omega(\theta)$ is the cross-correlation function measured from our \textcolor{red}{dataset}, $\omega_\mathrm{model}(\theta)$ is the cross-correlation function as defined by Equation \ref{eq:model1}, and $C_{ij}^{-1}$ is the inverse of covariance matrix given by the Equation \ref{eq:cov_matrix_boot}. We list the best-fit parameters in the legends of Figure \ref{fig:cross-corr}. The best-fit models are essentially flat straight lines. We do not report the uncertainties as the system is unbounded i.e., infinite combinations of $A_\mathrm{\omega}$ and $\beta$ can yield a flat line on the horizontal axis beyond a few arcsecond scale.

We find that for both GOODS-S and XDF samples, the best-fit parameters follow Equation \ref{eq:model2}, suggesting that any contamination that may exist in this sample is too small to be measurable by the cross-correlation analysis.
Nevertheless, we know that at least one of 191 sources in the GOODS-S catalog is spectroscopically confirmed to be a low-redshift source, and therefore the minimum contamination fraction (i.e., the ratio of the number of contaminants to the number objects identified as high-redshift galaxies) in the GOODS-S field is $f_\mathrm{cont}\geq 0.5 \%$. 
This result suggests to consider what level of contamination would introduce a measurable 
signal using this technique. We answer this question in the next Section, using mock observations from a cosmological simulation.



\section{Modeling the Cross-Correlation Function}
In this Section, we model the cross-correlation function as a function of contamination level.
To do so, we perform the same analysis in Section \ref{sec:cross-correlation} on the mock catalogs generated from \textit{The Next Generation Illustris} simulations \citep[IllustrisTNG,][]{springel2018,nelson2018,pillepich2018,naiman2018,marinacci2018} and use a Monte Carlo method to randomly select contaminants from interlopers.

\subsection{Illustris Mock Catalog}
The IllustrisTNG simulation suite is a collection of large-volume cosmological magnetohydrodynamical simulations that model galaxy formation, galaxy evolution, and large-scale structure formation within the $\Lambda$ cold dark matter paradigm. It is the follow-up project of the Illustris simulation series \citep{genel2014, vogelsberger2014a, vogelsberger2014b, nelson2015, sijacki2015}. Similarly to Illustris, IllustrisTNG is run using the quasi-Lagrangian moving-mesh code \textsc{arepo} \citep{Springel10}, which combines aspects of smooth particle-based hydrodynamical simulations with adaptive mesh-based simulations, in order to avoid numerical issues that each of these other methods possess \citep{Vogelsberger+13}. Subgrid physical prescriptions are used to model a large variety of astrophysical processes that are relevant for galaxy formation and evolution, including stochastic star formation, black hole formation and growth, stellar and AGN feedback, metal enrichment and cooling, and cosmic magnetic fields. The complete description of the TNG galaxy formation model is presented in the two TNG methods papers \citep{weinberger2017, pillepich2017}. 

In this study, we use the TNG300-1 simulation from the IllustrisTNG simulation suite, with a volume of 302.6 cMpc$^3$ and a mass resolution of $~10^7 M_\odot$ per baryonic particle. From this simulation, we generate multiple mock catalogs of high-redshift and intermediate-redshift galaxies. First, we download Snapshots 14 and 43 of this simulation from the TNG public database\footnote{\url{https://www.tng-project.org/data/downloads/TNG300-1/}}, which correspond to redshift $z=1.30$ and $z=5.85$, respectively.
Next, we select galaxies within 170 non-overlapping cutouts from each snapshot to generate position and photometry catalogs.
The dimensions of each cutout box at $z=1.30$ is ($13.34\times8.33\times151.90$) cMpc, and ($17.44\times27.91\times165.30$) cMpc at $z=5.85$.
These box sizes were chosen to be equivalent to a projected $0.12\times0.2$ degree sky survey in the observer frame, which is approximately the size of the GOODS-S area. In Snapshot 14 at $z=5.85$, all sub-boxes are oriented such that their long sides align with the Z-axis of the TNG300-1 simulation, with the central X- and Y- positions evenly drawn from a $10 \times 17$ grid that avoids the edges of the simulation volume. In Snapshot 43 at $z=1.30$, we instead orient the cutout boxes so that the long side aligns with the X-axis of the TNG300-1 simulation, and evenly sample the Y- and Z- positions from a $10 \times 17$ grid that avoids both the edges of the simulation, and the area of the simulation volume from which the cutout boxes at $z=5.85$ were drawn. This was done to avoid spurious correlations between galaxy clusters observed at $z=5.85$ and their own progenitors at $z=1.30$.
The generated catalogs contain the position, stellar mass, and photometry of the sources in rest-frame $U$, $B$, $V$, $K$, $g$, $r$, $i$, and $z$ bands. 


We convert the $i$ absolute magnitude of galaxies at redshift $z=1.30$ into apparent magnitude at wavelength $\lambda=17204$ \AA.
\begin{equation}
    m = M + 5 \log\left(\frac{D_L}{10\ \mathrm{pc}}\right)  -2.5\log(1+z),
\end{equation}
where $M$ is the absolute magnitude in emitted frame, $m$ is the apparent magnitude in observer frame, ${D_L}$ is the luminosity distance, and $z$ is the source's redshift. 
Similarly, we convert the $U$ magnitude of galaxies at $z=5.85$ into apparent magnitude at $\lambda=25688$ \AA. These wavelengths are the closest wavelengths to the detection bands of the \cite{merlin2021} and \cite{bouwens2021} catalogs where the photometry information is available. 
To simulate with a condition close to the current observation limit, we only select galaxies with an apparent magnitude up to 28.5 for both $z=1.30$ and $z=5.85$ galaxies.

To ensure that the mock fields from the simulations match the observation geometrically, we first rotate the mock field to match the position angle of the observation. We then apply the same field of view to ensure that the edges of the mock field have the same shape as those of the observation. Lastly, we apply the segmentation map generated by \texttt{SExtractor} \citep{bertin1996} to cut out the mock galaxies that would have been blocked by foreground galaxies in the real observation. 
As a sanity check, we measure the average angular correlation function of the simulated galaxies that survive the geometry cut above at both $z=1.3$ and $z=5.85$. The average angular correlation functions of all simulated fields are consistent with the observed angular correlation functions of $z\sim1.3$ and $z\sim6$ galaxies within 1$\sigma$ error.

\subsection{Monte Carlo Simulation}
\label{sec:monte-carlo-sim}
\begin{figure*}
    \centering
    \includegraphics[width=0.9\textwidth]{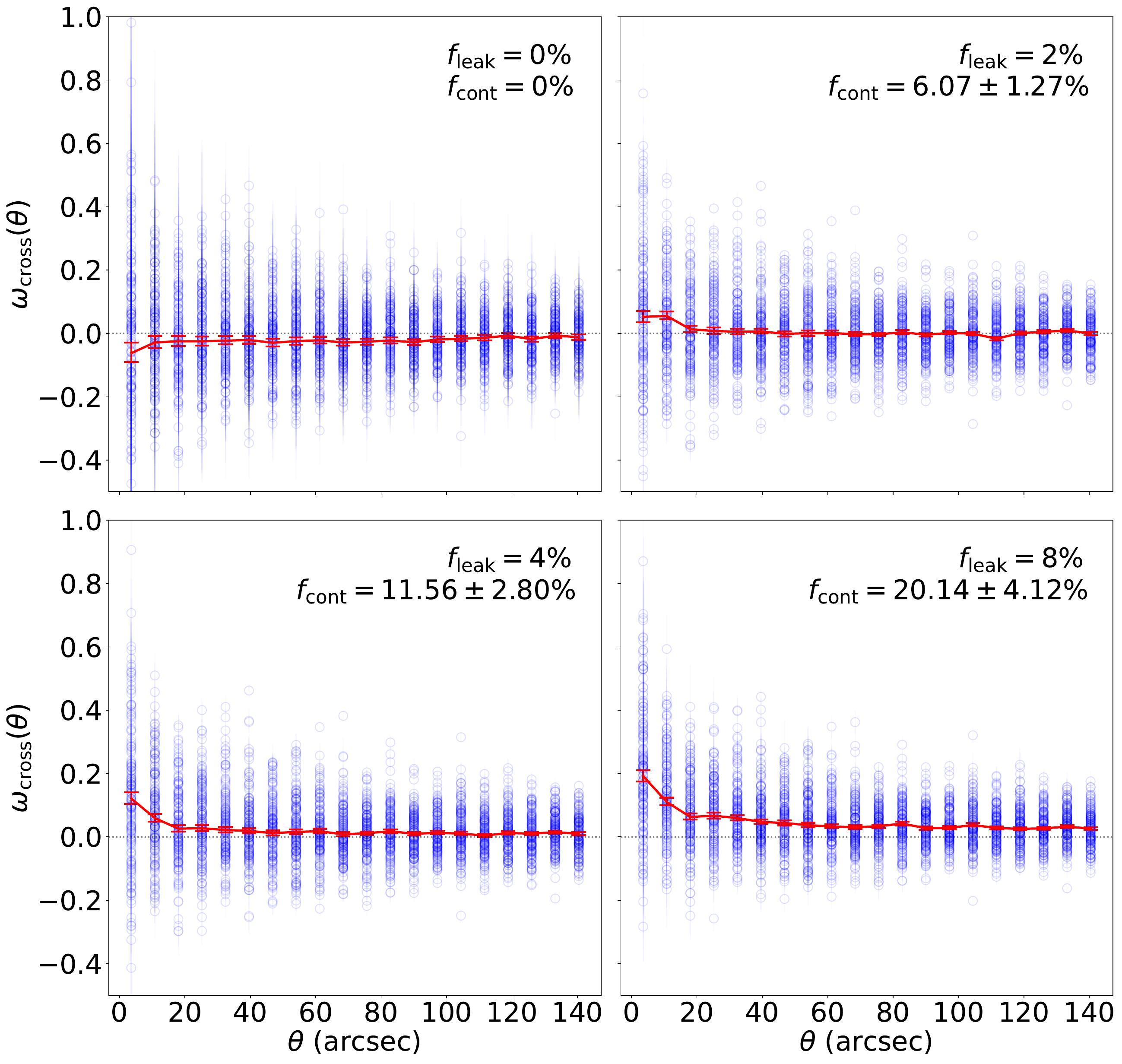}
    \caption{Cross-correlation function of galaxies at redshift $z\sim1.30$ and $z\sim5.85$ based on Illustris mock catalogs. The blue circles represent each data point generated from the Monte Carlo simulation. The red line is the mean and its standard error. Using a Monte Carlo simulation, we adjust the contamination level by increasing the leakage fraction ($f_\mathrm{leak}$). As we change $f_\mathrm{leak}$ from 0 to 8\%, the cross-correlation signal increase proportionally.}
    \label{fig:Illustris}
\end{figure*}

We define sources with apparent magnitude $\geq 24.0$ from the mock catalog of $z=1.30$ as faint intermediate-redshift galaxies.
In our simulation, these sources can contaminate the LBG sample.
We introduce a leakage fraction $f_\mathrm{leak}$ as the probability that a faint $z=1.30$ galaxy will be misidentified as a $z=5.85$ galaxy and become a contaminant. 
We perform a Monte Carlo simulation to study how the cross-correlation function changes as a function of the leakage fraction. 
We conduct the simulation for $f_\mathrm{leak}=0\%-10\%$ with a step size of $1\%$.
For $f_\mathrm{leak}=0\%$, we calculate the cross-correlation function in the same way as the previous section. 
We use the bootstrap resampling method to derive the mean and the error of the cross-correlation function.
For $f_\mathrm{leak}>0\%$, we estimate the uncertainty using the Monte Carlo method. For each sub-box, we repeat the process of assigning different $z\sim1.3$ galaxies as contaminants according to the leakage fraction and remeasuring the cross-correlation function ten times for each sub-box. The means and uncertainties for these individual subboxes are shown as blue circles in Figure \ref{fig:Illustris}.
The weighted mean of all 170 sub-boxes are shown as red lines in the same figure.
To ease the interpretation, we also convert the leakage fraction into the contamination fraction ($f_\mathrm{cont}$) from each simulation. The average value for $f_\mathrm{cont}$ is displayed for each panel in Figure \ref{fig:Illustris}.
Based on Figure \ref{fig:Illustris},
increasing the leakage fraction will increase the clustering signal in the cross-correlation function. Thus, a clustering signal in the cross-correlation function indicates contamination, in agreement with our hypothesisand the results from previous studies \citep{gebhardt2019,addison2019,awan2020,farrow2021}.

\begin{figure*}
    \centering
    \includegraphics[width=0.98\textwidth]{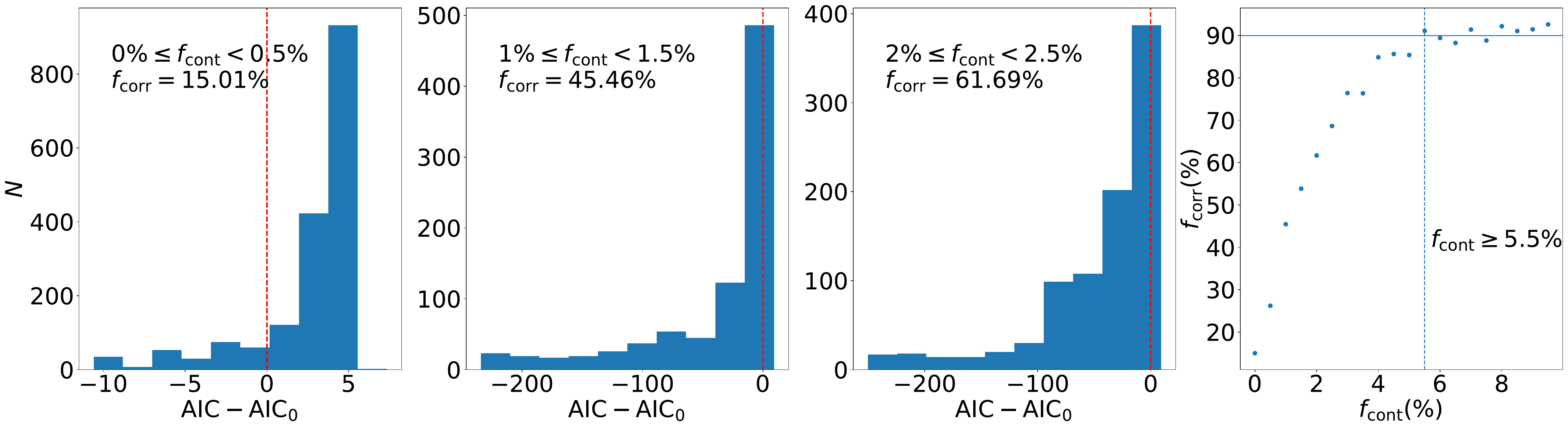}
    \caption{\emph{First three left panels}: Histograms of $\mathrm{AIC-AIC_0}$ for three bins of $f_\mathrm{cont}=0\%-0.5\%,1\%-1.5\%,2\%-2.5\%$. Negative value (left side of dashed vertical line) indicating that the simulation shows a cross-correlation signal. \emph{Right panel}: Fraction of simulation showing a cross-correlation signal ($f_\mathrm{corr}$) as a function of contamination fraction. For contamination fractions greater than $f_\mathrm{cont}=5.5\%$ (dashed vertical line), most simulations show a cross-correlation signal, as indicated by $f_\mathrm{corr}\sim90\%$ (horizontal line).}
    \label{fig:Illustris_2}
\end{figure*}

To estimate at what level of contamination we will see a significant cross-correlation signal, we conduct a statistical test for each combination of simulation box. 
We consider a hypothesis test where the null hypothesis ($H_0$) is that Equation \ref{eq:model1} and Equation \ref{eq:model2} fit the data equally well -- i.e. that there is no cross-correlation between the sample of high-redshift galaxies and the sample at the interloper redshift. 
The alternative hypothesis ($H_1$) is that Equation \ref{eq:model1} fits the data significantly better than Equation \ref{eq:model2}. Thus, we should use Equation \ref{eq:model1} where the measured cross-correlation function can be parameterized as a power-law function.
To determine which model is preferred (Equation \ref{eq:model1} or Equation \ref{eq:model2}), we use the Akaike information criterion (AIC, \citealt{akaike1974}), taking the number of parameters into account.
The model with a smaller AIC value is preferred.
For each model, the AIC value is given by:
\begin{equation}
    \mathrm{AIC}=2k-2\ln(\mathcal{L}),
    \label{eq:AIC}
\end{equation}
where $k$ is the number of free parameters, and $\mathcal{L}$ is the likelihood of the model given by: 
\begin{equation}
\begin{aligned}
    \mathcal{L} & = & \prod_{i,j}\frac{1}{2\pi^{p/2}|C_{ij}|^{1/2}}\exp{\Bigl\{-\frac{1}{2} \left[\omega(\theta_i)-\omega_\mathrm{model}(\theta_i)\right]^\mathrm{T}} C_{ij}^{-1}\\
    & & \left[\omega(\theta_j)-\omega_\mathrm{model}(\theta_j)\right]\Bigr\},
    \label{eq:likelihood}
\end{aligned}
\end{equation}
where $p$ is the number of bins in $\theta$, $|C_{ij}|$ is the determinant of the covariance matrix, $\omega(\theta)$ is the measured cross-correlation function, and $\omega_\mathrm{model}(\theta)$ is the modeled cross-correlation function based on Equation \ref{eq:model1} or Equation \ref{eq:model2}. 
Due to the small number of simulated galaxies, the errors in cross-correlation functions are dominated by the Poisson noise. Therefore, we use only the diagonal elements in the covariance matrix, in the same manner as \cite{zheng2007} and \cite{harikane2016}. By using the GOODS-S and XDF observational data in Section \ref{subsec:cross_corr_obs}, we have tested that the use of the off-diagonal elements in covariance matrix does not change the conclusion of the best-fit model.
Using a fixed integral constraint (IC) derived from all of the simulations, we calculate $\mathrm{AIC-AIC_0}$ -- that is, the difference between the AIC value of model (\ref{eq:model1}) and the AIC value of model (\ref{eq:model2}). 
If this value is negative, then the simulation produces a significant cross-correlation signal. 
We calculate how many of the simulations produce significant correlations as a function of contamination fraction. 
We generate a histogram of $\mathrm{AIC-AIC_0}$ for each bin of 0.5\% contamination fraction. 
We show our result in Figure \ref{fig:Illustris_2}. 
For $f_\mathrm{cont}\geq5.5\%$, $\sim90\%$ of simulations consistently show cross-correlation signal.
Therefore, we conclude that the level of contamination in the GOODS-S field is less than 5.5\% (at 90\% confidence).



\begin{figure*}
    \centering
    \includegraphics[width=0.9\textwidth]{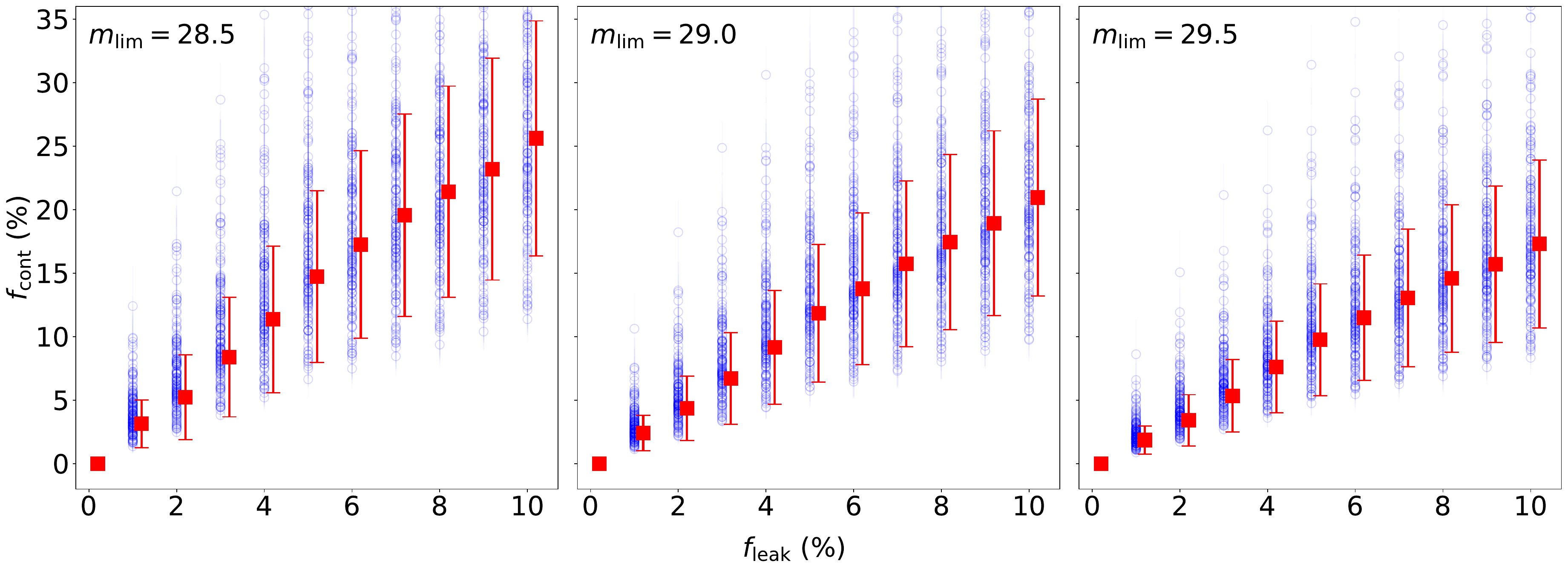}
    \caption{Contamination fraction as a function of leakage fraction for galaxies at redshift $z\sim1.30$ and $z\sim5.85$ based on Illustris mock catalogs using three different magnitude cuts. The blue circles represent each data point generated from the Monte Carlo simulation. The red square is the mean and its standard deviation (slightly shifted to the right for visual clarity). By adjusting the limiting magnitude of our sample in the catalogs, we find that for deeper field, the contamination fraction ($f_\mathrm{cont}$) is lower for the same value of leakage fraction ($f_\mathrm{leak}$). Therefore, contamination can be minimised with deeper observations.} 
    \label{fig:Illustris_3}
\end{figure*}

To test how the contamination level depends on the depth of the survey, we repeat our Monte Carlo analysis process using two deeper limiting magnitudes of 29.0 and 29.5.
We present the results of this experiment in Figure \ref{fig:Illustris_3}.
As we use the fainter magnitude cut, the error bars of the cross-correlation become smaller.
The clustering signal for the same value of leakage fraction also becomes smaller, indicating a lower contamination fraction.

We conclude that the contamination level depends on the depth of the survey. This result can be explained as a consequence of the steepening of high-redshift galaxy UV Luminosity Function towards the faint-end. As the depth of a survey is increased, the number of actual high-redshift galaxies increases more rapidly than the number of intermediate-redshift interlopers.

\subsection{Contamination Fraction Calculation based on Previous Literature}
\label{sec:f_cont_awan}
\cite{awan2020} introduces a formalism that uses the observed cross-correlation function in the contaminated sample to estimate the true cross-correlation function.
Based on their work, the observed cross correlation function is contributed by four types of pairings. In our context, the four parings are: 
(1) between true $z=1.3$ galaxies and true $z=6$ galaxies, 
(2) true $z=1.3$ galaxies and observed $z=6$ galaxies that are actually at $z=1.3$, 
(3) true $z=6$ galaxies and observed $z=1.3$ galaxies that are actually at $z=6$, and
(4) observed $z=1.3$ galaxies that are actually at $z=6$ and observed $z=6$ galaxies that are actually at $z=1.3$.
The final observed cross correlation function is:
\begin{equation}
\begin{aligned}
     \omega_\mathrm{cross}^\mathrm{obs}(\theta) = & f_\mathrm{z\sim1.3}^\mathrm{true}f_\mathrm{z\sim6}^\mathrm{true}\omega_\mathrm{cross}^\mathrm{true}(\theta) + \\
     & f_\mathrm{z\sim1.3}^\mathrm{true}f_\mathrm{z\sim6}^\mathrm{cont}\omega_{z\sim1.3}^\mathrm{true}(\theta) + \\
     & f_\mathrm{z\sim1.3}^\mathrm{cont}f_\mathrm{z\sim6}^\mathrm{true}\omega_{z\sim6}^\mathrm{true}(\theta) + \\
     & f_\mathrm{z\sim6}^\mathrm{cont}f_\mathrm{z\sim1.3}^\mathrm{cont}\omega_\mathrm{cross}^\mathrm{true}(\theta),
     \label{eq:awan}
\end{aligned}
\end{equation}
where $\omega_\mathrm{cross}^\mathrm{obs}(\theta)$ is the observed cross-correlation function, $\omega_\mathrm{cross}^\mathrm{true}(\theta)$ is the true cross-correlation function, $\omega_{z\sim1.3}^\mathrm{true}(\theta)$ is the true angular correlation function (ACF) for galaxies at $z\sim1.3$, $\omega_{z\sim6}^\mathrm{true}(\theta)$ is the true ACF for galaxies at $z\sim6$, $f_\mathrm{z\sim1.3}^\mathrm{true}$ is the fraction of galaxies at $z\sim1.3$ that are not contaminant from $z\sim6$ galaxies, $f_\mathrm{z\sim6}^\mathrm{true}$ is the fraction of galaxies at $z\sim6$ that are not contaminant from $z\sim1.3$ galaxies, $f_\mathrm{z\sim1.3}^\mathrm{cont}$ is contamination fraction in $z\sim1.3$ galaxies sample, and $f_\mathrm{z\sim6}^\mathrm{cont}$ is the contamination fraction in $z\sim6$ galaxies sample.

In our case, we assumed that the galaxies sample at lower redshift is not contaminated ($f_\mathrm{z\sim1.3}^\mathrm{cont}=0$, $f_\mathrm{z\sim1.3}^\mathrm{true}=1$) and the true cross-correlation function should be zero ($\omega_\mathrm{cross}^\mathrm{true}(\theta)=0$). Therefore, the first, third, and fourth term in Equation \ref{eq:awan} vanish, and the expression can be simplified into:
\begin{equation}
     \omega_\mathrm{cross,obs}(\theta) = f_\mathrm{cont}\omega_{z\sim1.3}(\theta),
     \label{eq:awan_simplified}
\end{equation}
where $\omega_\mathrm{cross,obs}(\theta)$, $f_\mathrm{cont}$, and $\omega_{z\sim1.3}(\theta)$ are the observed cross-correlation function, the contamination fraction in high-redshift galaxies sample, and the ACF of intermediate-redshift interlopers, respectively.

We estimate the angular correlation function of our $z\sim1.3$ sample following the power-law form: $\omega_{z\sim1.3}(\theta)=A_\omega\theta^{-\beta}$. The best-fit parameters are $A_\omega = 0.66\pm0.18$ and $\beta = 0.45\pm0.19$ for the GOODS-S field and $A_\omega = 0.48^{+0.59}_{-0.48}$ and $\beta = 0.81\pm0.79$ for the XDF field. We conduct the $\chi^2$ minimization method to find the best contamination fraction ($f_\mathrm{cont}$) based on the observed cross-correlation function calculated in Section \ref{subsec:cross_corr_obs}. Our results for the contamination fraction are $f_\mathrm{cont,GOODS-S} = 0.00^{+2.87}_{-0.00} \%$ for the GOODS-S field and $f_\mathrm{cont,XDF} = 0.00^{+6.76}_{-0.00} \%$ for the XDF field. These results agree with the estimations from our Monte Carlo simulation in Section \ref{sec:monte-carlo-sim}, i.e. the contamination level should be less than $5.5\%$ if we do not detect any cross-correlation signal.




\section{Number Count Analysis}
 The spatial cross-correlation analysis is most appropriate for surveys with a large contiguous field of view. Still, it may not apply to surveys with several pencil beam observations, such as random-pointing multiple field surveys. In this Section, rather than using the spatial cross-correlation analysis presented above, we explore an alternative method to constrain the contamination by using a simple correlation between the number of the targeted population and the number of possible interloper populations. 
 This method is based on the theoretical paper by \cite{robertson2010} who measures clustering of high-redshift galaxies based on counts-in-cell analysis.
 The idea is that there should be no correlation in the number counts across the observed fields unless contamination exists. Based on available data in the literature, we test this principle on the BoRG data set, setting $z\sim8$ galaxies as our high-redshift galaxy sample. The corresponding interloper redshift is $z\sim2$. We describe the data set in Section \ref{sec:borgdata}. The correlation analysis is in Section \ref{sec:correlation_borg}. We then discuss the simulation in Section \ref{sec:sim_borg} and interpret the results in Section \ref{sec:result_borg}.

\subsection{Data Set}\label{sec:borgdata}
We use samples of $z\sim2$ and $z\sim8$ galaxies from the Brightest of Reionizing Galaxies (BoRG) survey \citep{trenti2011}. 
The BoRG survey is a pure-parallel program on the Hubble Space Telescope focused on finding bright galaxy candidates at redshift $z \gtrsim 7$ using the Lyman break technique \citep{steidel1996}. Specifically, we take $z\sim8$ galaxies from the catalogs provided by \citet{bradley2012} and \citet{schmidt2014} and $z\sim2$ galaxies from the catalogs \citet[hereafter \citetalias{cameron2019}]{cameron2019}.

Both catalogs are based on the first generation of the BoRG data release, which consists of 71 random independent pointings taken with three different near-infrared filters (WFC3/IR F098M, F125W, and F160W) and one optical filter (either WFC3/UVIS F606W or F600LP). \citetalias{cameron2019} discards two fields because they affected by star overdensity and significant Galactic dust-reddening. Hence, the number of overlapping search fields between the two catalogs is 69 fields. Due to the nature of the pure-parallel survey, each field has a different exposure time, which leads to $5\sigma$ limiting magnitude in F125W ranging between $25.6 - 27.5$. 

The $z\sim2$ galaxies in \citetalias{cameron2019} were detected in $H_{160}$ band and selected using $Y_{098} - H_{160} > 1.5$ cut. Photometric redshift estimates were obtained with the Bayesian photometric redshift code \texttt{BPZ} \citep{benitez2000,benitez2004,coe2006}. The redshift range of \citetalias{cameron2019} final sample is $1.5 < z < 2.5$. The catalog consists of 490 galaxies and is expected to be highly complete and not contaminated up to $m_{\mathrm{AB,}H}=24.5$.
On the other hand, the $z\sim8$ galaxies in \citet{bradley2012} and \citet{schmidt2014} were detected in $J_{125}$ band and were selected using Lyman break technique that includes objects in the redshift range $7.5 \lesssim z \lesssim 8.5$. The catalog consists of 42 $z\sim8$ galaxies with F125W magnitude ranging from 25.50 to 27.60.


\subsection{Correlation between Number Counts of Galaxies at Two Redshifts in BoRG Data}
\label{sec:correlation_borg}

\begin{figure*}
	\centering
	\includegraphics[width=\textwidth]{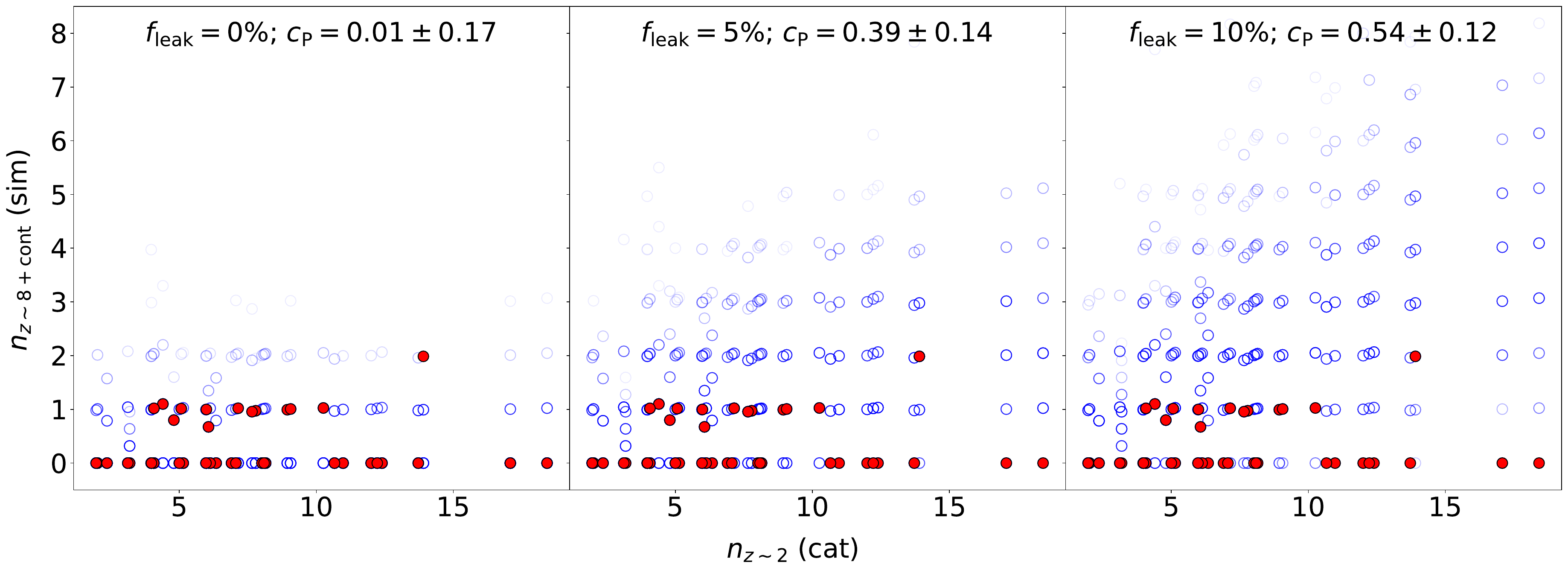}
    \caption{Comparing the number of galaxies observed at intermediate redshift $(z\sim 2)$ to the number of galaxies at high redshift ($z\sim8$) for all 39 fields of the BoRG survey studied in this paper. The filled red circles in every panel show the values taken from the $z \sim 2$ catalogue of \citetalias{cameron2019} and the $z\sim 8$ catalogues of \citet{bradley2012} and \citet{schmidt2014}.
    The blue circles are generated from our Monte Carlo simulation, using three different values of the leakage fraction to compute the number count of $z\sim8$ galaxies plus contaminants. The transparency of blue circles represents the frequency of Monte Carlo draws (more transparent means less occurrence). The Pearson correlation coefficient $c_\mathrm{P}$ is also shown for each case of leakage fraction. All the presented number counts are rescaled to a median field area of $4.40\ \mathrm{arcmin}^2$.}
    \label{fig:BoRG_data}
\end{figure*}

Due to the different depths for each field in our sample, we may introduce an artificial correlation. 
Deeper fields may have a higher number count of $z\sim2$ and $z\sim8$ galaxies than shallow fields. The median depth among all the fields is 26.75 mag in F125W band.
We therefore remove all fields with limiting magnitude fainter than 26.75 mag and all galaxies with F125W fainter than 26.75 to get a sample of $z\sim8$ sources that have the same completeness across magnitude bins up to F125W$=26.75$ mag. The completeness of these $z\sim8$ galaxies is approximately $60\%$ \citep{trenti2012}. Our final sample consists of 39 fields with 306 $z\sim2$ galaxies and 14 $z\sim8$ galaxies. The number counts between the $z\sim8$ and $z\sim2$ galaxies in each field are plotted as red circles in Figure \ref{fig:BoRG_data}.

To quantify the correlation between the number of $z\sim2$ and $z\sim8$ galaxies, we measure the Pearson correlation coefficient. We use Fisher's transformation to estimate the confidence interval of Pearson correlation coefficient (a value of 1 indicates a perfect correlation, while a value of 0 indicates no correlation). Our Pearson correlation coefficient is equal to $0.05\pm0.17$. Although it is positive at face value, it is consistent with zero within $1\sigma$. Regardless, we proceed to measure the corresponding contamination fraction using a Monte Carlo simulation in the following Section.

\subsection{Simulations}
\label{sec:sim_borg}
To estimate the contamination fraction in the BoRG sample, we perform a Monte Carlo simulation following the methodology outlined in Section \ref{sec:monte-carlo-sim}. 
We first calculate the expected number of faint $z\sim2$ galaxies and the expected number of $z\sim8$ galaxies specific to each observed field. Because the sample of $z\sim2$ galaxies is pure and complete up to F125W$=24.5$, the interlopers are likely to come from the population with apparent magnitudes fainter than 24.5. We define all $z\sim2$ galaxies with a magnitude between 24.5 to 26.75 in the F125W band as faint $z\sim2$ galaxies. These galaxies are not included in the \citetalias{cameron2019} catalogue. 

We estimate the expected number count of faint $z\sim2$ galaxies in each field of the BoRG survey using the luminosity function of $z\sim2$ galaxies from \citet{marchesini+12}. Based on this luminosity function, we calculate the ratio of the number of faint galaxies with $24.25<$ F125W $<26.75$ to the number of bright galaxies with F125W$\leq24.5$. We then normalize (multiply) the ratio with the observed number count of bright $z\sim2$ galaxies from the catalog to estimate the expected number count of faint $z\sim2$ galaxies in the field.
We also calculate the expected number count of $z\sim8$ galaxies (i.e., the number count of real $z\sim8$ galaxies) based on the UV luminosity function with the Schechter parameters from \citet{schmidt2014} and the assumption that the detection is 60\% complete. 
We present these expected number counts for each field of the BORG survey that we study in Table \ref{tab:numbercount}.

\begin{table}
	\centering
	\caption{The number ($n^\mathrm{cat}$) of $z\sim2$ and $z\sim8$ galaxies within each of the 39 fields from the BoRG catalogs considered in this analysis. We also tabulate the expected number count of faint $z\sim2$ galaxies, and the intrinsic number count of $z\sim8$ galaxies within each field predicted from galaxy luminosity functions ($n^\mathrm{LF}$).}
	\label{tab:numbercount}
	\begin{tabular}{lccccc} 
    \hline
    Field name & Area & $n^\mathrm{cat}_{z\sim2}$ & $n^\mathrm{LF}_{\mathrm{faint}, z\sim2}$ & $n^\mathrm{cat}_{z\sim8}$ & $n^\mathrm{LF}_{\mathrm{int}, z\sim8}$\\
    \hline
0110--0224& 13.81&   10& 	 20& 		 0& 		 0.83\\
0228--4102& 4.43&  	 4& 	 8& 		 0& 		 0.27\\
0436--5259& 4.33&  	 4& 	 8& 		 0& 		 0.26\\
0439--5317& 4.28& 	 5& 	 10& 		 0& 		 0.26\\
0440--5244& 4.34& 	 5& 	 10& 		 1& 		 0.26\\
0553--6405& 4.00& 	 4& 	 8& 		 1& 		 0.24\\
0751+2917&  4.52&	 8& 	 16& 		 1& 		 0.27\\
0846+7654&  4.41&	 11& 	 22& 		 0& 		 0.26\\
0906+0255&  4.39&	 8& 	 16& 		 0& 		 0.26\\
0914+2822&  4.40&	 12& 	 24& 		 0& 		 0.26\\
0952+5304&  4.42&	 4& 	 8& 		 0& 		 0.26\\
1010+3001&  4.54&	 11& 	 22& 		 0& 		 0.27\\
1031+5052&  5.55&	 8& 	 16& 		 0& 		 0.33\\
1033+5051&  5.50&	 6& 	 12& 		 1& 		 0.33\\
1051+3359&  4.26&	 12& 	 24& 		 0& 		 0.26\\
1059+0519&  4.43&	 9& 	 18& 		 1& 		 0.27\\
1103--2330& 4.37&	 9& 	 18& 		 1& 		 0.26\\
1111+5545&  4.31&	 6& 	 12& 		 0& 		 0.26\\
1118--1858& 4.23&	 3& 	 6& 		 0& 		 0.25\\
1119+4026&  4.46&	 7& 	 14& 		 0& 		 0.27\\
1131+3114&  4.41&	 6& 	 12& 		 1& 		 0.26\\
1152+5441&  4.40&	 5& 	 10& 		 0& 		 0.26\\
1209+4543&  4.42&	 6& 	 12& 		 0& 		 0.26\\
1242+5716&  4.29&	 10& 	 20& 		 1& 		 0.26\\
1341+4123&  4.36&	 7& 	 14& 		 0& 		 0.26\\
1358+4326&  4.49&	 14& 	 28& 		 0& 		 0.27\\
1358+4334&  4.32&	 8& 	 16& 		 0& 		 0.26\\
1408+5503&  4.32&	 4& 	 8& 		 1& 		 0.26\\
1416+1638&  4.38&	 17& 	 34& 		 0& 		 0.26\\
1429--0331& 4.35&	 8& 	 16& 		 0& 		 0.26\\
1437+5043&  6.53&	 9& 	 18& 		 1& 		 0.39\\
1459+7146&  4.32&	 12& 	 24& 		 0& 		 0.26\\
1510+1115&  4.43&	 14& 	 28& 		 2& 		 0.27\\
1555+1108&  4.31&	 7& 	 14& 		 1& 		 0.26\\
1632+3733&  4.37&	 2& 	 4& 		 0& 		 0.26\\
2203+1851&  4.60&	 8& 	 16& 		 1& 		 0.28\\
2313--2243& 5.59&	 3& 	 6& 		 0& 		 0.33\\
2345+0054&  4.48&	 2& 	 4& 		 0& 		 0.27\\
2351--4332& 4.30&	 18& 	 37& 		 0& 		 0.26\\
    \hline
	\end{tabular}
\end{table}

Finally, we perform a Monte Carlo simulation to study how the correlation between the number of $z\sim2$ and $z\sim8$ galaxies correlation changes with the leakage fraction. We draw a random number count of faint $z\sim2$ galaxies and real $z\sim8$ galaxies following the Poisson distribution $f(k;\lambda)=\lambda^ke^{-\lambda}/k!$, where $\lambda$ is their expected numbers from the luminosity functions. Then, the number of contaminants is estimated based on the simulated number of faint $z\sim2$ galaxies and the value of leakage fraction. The number of observed $z\sim8$ galaxies is the number of true $z\sim8$ galaxies plus the number of contaminants. As done with the real data, we calculate the Pearson correlation coefficient $c_\mathrm{P}$ between the observed number of bright $z\sim2$ galaxies (which is the same as that of the real data) and the observed number of $z\sim8$ galaxies. We conduct the procedure for $f_\mathrm{leak} = 0\% - 10\%$ with a step size of $0.2\%$ and repeat the simulation 100 times at each value of the leakage fraction.
 
\subsection{Results and Discussion}
\label{sec:result_borg}
\begin{figure}
    \centering
	\includegraphics[width=0.48\textwidth]{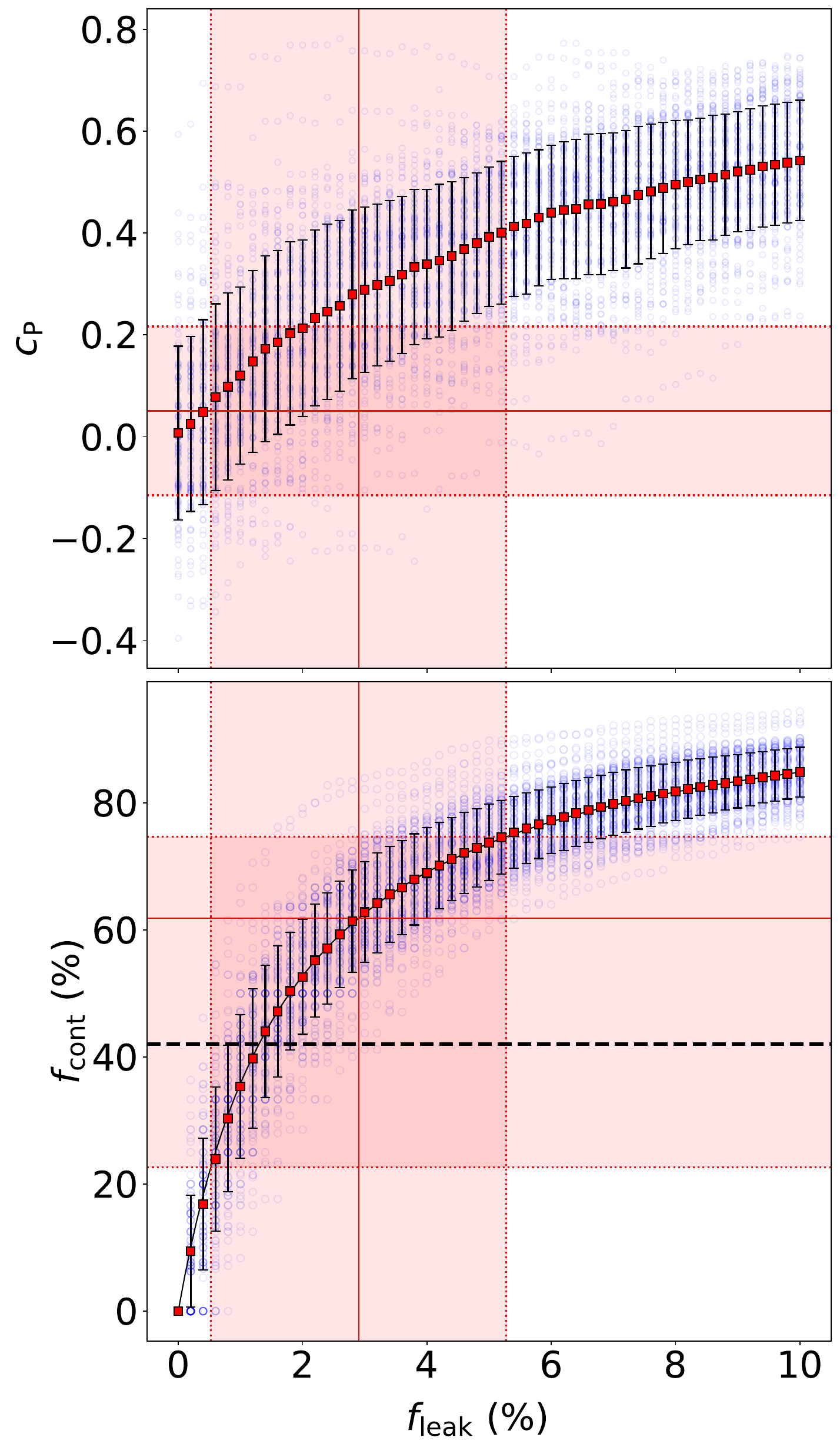}
    \caption{\emph{Top panel}: Pearson correlation coefficient as a function of leakage fraction. The blue circles represent each data point generated from the Monte Carlo simulation. The red square is the mean and its standard deviation. The red horizontal line and its shaded region are Pearson correlation coefficient based on catalog and its $1\sigma$ error, respectively. The red vertical line and its shaded region are the estimated leakage fraction and its $1\sigma$ error, respectively. \emph{Bottom panel}: Contamination fraction as a function of leakage fraction. The red vertical line and its shaded region are same as the top panel. The red horizontal line and its shaded region are our contamination fraction estimates and the $1\sigma$ error, respectively. The black dashed horizontal line is the previous estimate from \citet{bradley2012}.}
    \label{fig:BoRG_result_2}
\end{figure}

We present how the Monte Carlo simulation works in Figure \ref{fig:BoRG_data}. Blue circles show the simulated number counts from all 100 simulations for three values of leakage fraction. 
As the leakage fraction is increased, the simulated number count of $z\sim8$ galaxies becomes more correlated with the observed number count of $z\sim2$ galaxies. This is due to more $z\sim2$ galaxies being misidentified as $z\sim8$ galaxies. From Figure \ref{fig:BoRG_data}, we see that the simulation most closely resembles the data (solid red circles) when the leakage fraction is close to zero.
The top panel in Figure \ref{fig:BoRG_result_2} shows the Pearson correlation coefficient derived from those simulations. As the leakage fraction increases, the number of contaminants increases. Consequently, the correlation between the number of $z\sim8$ samples and the number of $z\sim2$ samples becomes tighter (as indicated by the increasing value of $c_\mathrm{P}$).
To facilitate the interpretation, we present contamination fraction ($f_\mathrm{cont}$, a percentage of observed $z\sim8$ galaxies that are low-z interlopers) as a function of $f_\mathrm{leak}$ in the bottom panel of Figure \ref{fig:BoRG_result_2}. The contamination increases rapidly as a function for leakage fraction and plateaus at contamination fraction of $\sim80\%$. A mere leakage fraction of $1\%$ already corresponds to contamination fraction of $20-50\%$. 

To calculate the best-fit $f_\mathrm{leak}$ and contamination of the BoRG sample, we compute the weighted average of the leakage fraction from the simulation based on the correlation coefficient of the observations. 
For each blue data point in the upper row of Figure \ref{fig:BoRG_result_2} calculated from the simulation, we measure the weight by assuming a Gaussian distribution:
\begin{equation}
    g(x)=\frac{1}{\sigma\sqrt{2\pi}}\exp\left(-\frac{1}{2}\frac{(x-\mu)^2}{\sigma^2}\right),
	\label{eq:gaussian}
\end{equation}
where $x$ is the $c_\mathrm{P}$ value for each data point. In this Equation, $\mu$ and $\sigma$ are the Pearson correlation coefficient and its uncertainty estimated from the observation in Section \ref{sec:correlation_borg}, i.e.  $\mu=0.05$ and $\sigma=0.17$. Putting it into Equation \ref{eq:gaussian}, we calculate the weight of each data point generated from simulation. Then, we measure the weighted mean as our leakage fraction estimation. We calculate $f_\mathrm{leak}=2.90\pm2.38\%$, corresponds to $f_\mathrm{cont}=62^{+13}_{-39}\%$ (vertical red line in Figure \ref{fig:BoRG_result_2}). 
The uncertainty in our result is high because of the small size of our sample.
Our result is consistent with the previous estimate by \citealt{bradley2012} ($f_\mathrm{cont}=42\%$) and the follow-up observation of \citealt{livermore2018} ($f_\mathrm{cont}\sim50\%$) within the $1\sigma$ confidence interval. 
\cite{bradley2012} estimate the contamination fraction by degrading a F606W image of GOODS-ERS data that is deeper than BORG data. They then conduct the selection again on the degraded sample and compare to the original GOODS-ERS catalog. The contamination fraction can then be calculated by checking which low-redshift galaxy in the original catalog leaks into the catalog generated from the degraded images. 


\begin{figure*}
    \centering
	\includegraphics[width=\textwidth]{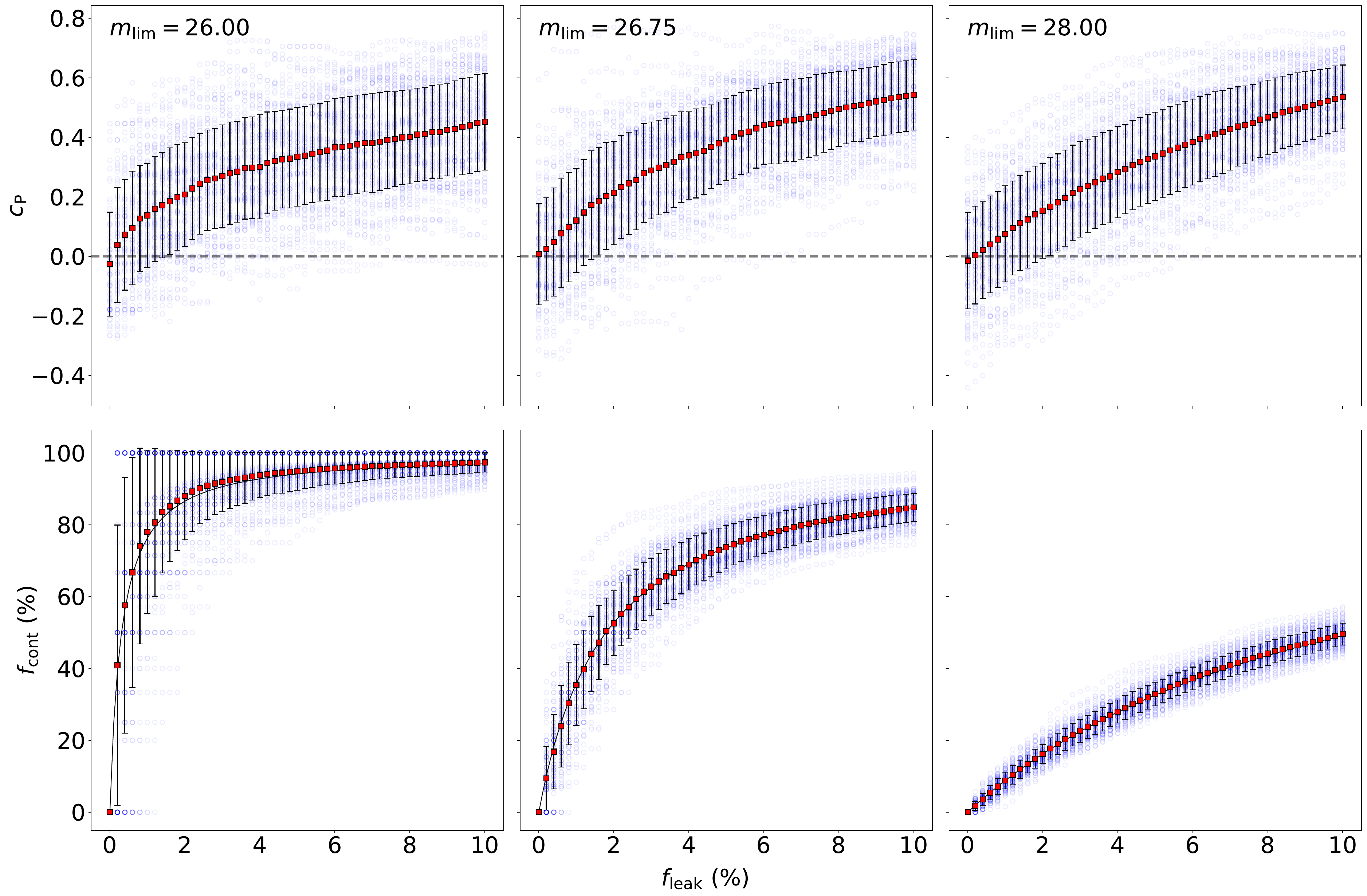}
    \caption{Same as Figure \ref{fig:BoRG_result_2}, but applied to two different limiting magnitude cases. \emph{Top panel}: Pearson correlation coefficient $c_\mathrm{P}$ as a function of leakage fraction $f_\mathrm{leak}$ for three different limiting magnitudes of 26.00, 26.75, 28.00 (left to right, respectively). The blue circles represent each data point generated from the Monte Carlo simulation. The red square is the mean and its $1\sigma$ error. \emph{Bottom panel}: Same as top panel, but the $y$-axis is the contamination fraction $f_\mathrm{cont}$. The black curve is the $f_\mathrm{cont}$ as a function of $f_\mathrm{leak}$ derived based on the luminosity function.}
    \label{fig:BoRG_result_1}
\end{figure*}

For comparison, we repeat the simulation assuming different limiting magnitudes at which the detection and the redshift determination for the low-$z$ population are complete, specifically at $m_\textrm{lim}=26.00$ and 28.00 (left and right columns of Figure \ref{fig:BoRG_result_1}). Unsurprisingly, our results indicate that deeper fields have less contamination. The faint-end of $z\sim8$ luminosity function is steeper than $z\sim2$ luminosity function. Therefore, it is expected that for a deeper observation, we will get more $z\sim8$ galaxies than $z\sim2$ galaxies. As the number count of contaminants follows the luminosity function of $z\sim2$ galaxies, $f_\mathrm{cont}$ will be lower for the same value of $f_\mathrm{leak}$.

\section{Summary}

We presented a novel analysis of contamination in Lyman-break galaxy samples at high redshift by studying the spatial correlation with the intermediate-redshift galaxies. We considered two methods based on the nature of high-redshift surveys: a large-contiguous-field survey and a multiple-field surveys.  As a demonstration of the two approaches, we investigated applications to the CANDELS GOODS-S and XDF survey, and to the BoRG random-pointing multiple field survey, respectively. 
We summarize our results as follows:
\begin{itemize}
    \item We carried out cross-correlation analysis based on the CANDELS data and performed statistical tests to quantify the contamination level in GOODS-S and XDF fields. Both fields show no significant cross-correlation signal between $z\sim6$ galaxies and lower redshift galaxies at the redshift of potential contaminants (i.e. $z\sim1.3$).
    \item Using the mock catalog generated based on IllustrisTNG simulation, we modelled the changes in the cross-correlation function as a function of the contamination fraction. As we increased the contamination, the cross-correlation signal becomes stronger. We estimated that for GOODS-S field, the contamination fraction is below $5.5\%$ at $90\%$ confidence level.
    \item Our analysis shows that for a deeper field, the contamination is lower than those with a shallow field. This can be explained based on the luminosity function. The luminosity function for high-redshift galaxies is steeper toward the faint-end compared to those of intermediate-redshift galaxies. Thus, the number of contaminants increases more slowly than the number of true high-redshift galaxies as the survey depth is increased.
    \item We applied a count-in-cell correlation analysis to a survey with a large number of independent lines of sight, using the relatively shallow BoRG dataset. We detected evidence of number counts correlation, with a quantitative analysis  estimating the contamination fraction for the BoRG $z\sim8$ sample to be $62^{+13}_{-39}\%$, consistent with the previous calculation by \cite{bradley2012} within $1\sigma$ confidence interval. The large error bar in our estimates is caused by the low average number of counts in each field, which gives rise to large Poisson fluctuations. 
\end{itemize}

Overall, we demonstrated the utility of our novel analysis as an independent check of contamination in Lyman-break galaxy samples.
We can apply our method to larger data sets expected to become available from upcoming JWST observations. For example, the number count analysis can be applied in the upcoming PANORAMIC Survey (A Pure Parallel Wide Area Legacy Imaging Survey at $1-5$ Micron, \citealt{williams2021}) and GO 3990: A NIRCam Pure-Parallel Imaging Survey of Galaxies Across the Universe \citep{morishita2023}. We expect those surveys will have larger sample sizes than the BoRG survey at $z\sim 8$, enabling higher precision measurements of the contamination fraction from count-in-cell correlation.

\section*{Acknowledgements}
We thank Daniel Joseph Farrow as the referee for useful suggestions and comments that have improved the manuscript. This research was supported by the Australian Research Council Centre of Excellence for All Sky Astrophysics in 3 Dimensions (ASTRO 3D), through project number CE170100013. MH acknowledges support from a Melbourne Research Scholarship. BM acknowledges support from an Australian Government Research Training Program (RTP) Scholarship. 
The authors thank Charlotte A. Mason for useful comments on the manuscript. MH also thanks Yuichi Harikane for advice during the preparation of the manuscript.

\section*{Data Availability}

The data underlying this article will be shared on reasonable request to the corresponding author.



\bibliographystyle{mnras}
\bibliography{mnras_main} 

\bsp	
\label{lastpage}
\end{document}